\documentclass[a4paper, 11pt]{article}

\usepackage{amssymb,amsmath}
\usepackage{commath}
\usepackage{bbm}
\usepackage{amsmath, amsthm, thm-restate}
\usepackage{MnSymbol}
\usepackage{mathrsfs}
\usepackage{tikz}
\usepackage{float}
\usepackage{braket}
\usepackage{graphicx}
\usepackage{color}
\usepackage{listings}
\usepackage[outercaption]{sidecap}  
\usepackage[top = 2.6cm, bottom = 2.6cm, left = 2.6cm, right = 2.6cm]{geometry} 
\usepackage[toc,page]{appendix}

\newtheorem{theorem}{Theorem}

\newtheorem{corollary}[theorem]{Corollary}
\newtheorem{lemma}[theorem]{Lemma}
\theoremstyle{definition}
\newtheorem{definition}[theorem]{Definition}
\theoremstyle{definition}
\newtheorem{observation}[theorem]{Observation}

\newcommand{\trace}{\text{Tr}}
\newcommand{\reals}{\mathbb{R}}
\newcommand{\bt}{\beta}
\newcommand{\bbeta}{\boldsymbol{\beta}}
\newcommand{\brho}{\boldsymbol{\rho}}
\newcommand{\vecsig}{\boldsymbol{\sigma}}
\newcommand{\veclam}{\boldsymbol{\lambda}}

\usepackage{hyperref}
\providecommand{\keywords}[1]
{
  \small	
  \textbf{Keywords:} #1
}
\title{The geometry of Bloch space in the context of quantum random access codes}
\makeatletter
\renewcommand\@date{{%
  \vspace{-\baselineskip}%
  \large\centering
  \begin{tabular}{@{}c@{}}
    Laura Man\v{c}inska  
  \end{tabular}%
  \quad  {} \quad
  \begin{tabular}{@{}c@{}}
    Sigurd A.~L.~Storgaard \footnote{qmt293@alumni.ku.dk}
  \end{tabular}

  \bigskip

  Department of Mathematical Sciences, University of Copenhagen, Denmark \par

  \bigskip

  \today
}}
\makeatother

\graphicspath{{figures/}}

\usepackage{blindtext}
\usepackage{subfiles}

\begin{document}

\definecolor{dkgreen}{rgb}{0,0.6,0}
\definecolor{gray}{rgb}{0.5,0.5,0.5}
\definecolor{mauve}{rgb}{0.58,0,0.82}

\newcommand{\lnote}[1]{\textcolor{magenta}{\small {\textbf{(Laura:} #1\textbf{) }}}}
\newcommand{\snote}[1]{\textcolor{red}{\small {\textbf{Sigurd:} #1\textbf{ }}}}
\newcommand{\psomit}[1]{\textcolor{blue}{\small {\textbf{Possibly omit:} #1\textbf{ }}}}
\newcommand{\Squestion}[1]{\textcolor{yellow}{\small {\textbf{(Question:} #1\textbf{) }}}}
\newcommand{\stcomp}[1]{{#1}^\complement} 


\maketitle

\begin{abstract}
We study the communication protocol known as a Quantum Random Access Code (QRAC) which encodes $n$ classical bits into $m$ qubits ($m<n$) with a probability of recovering any of the initial $n$ bits of at least $p>\tfrac{1}{2}$. Such a code is denoted by $(n,m,p)$-QRAC. If cooperation is allowed through a shared random string we call it a QRAC with shared randomness. We prove that for any $(n,m,p)$-QRAC with shared randomness the parameter $p$ is upper bounded by $ \tfrac{1}{2}+\tfrac{1}{2}\sqrt{\tfrac{2^{m-1}}{n}}$. For $m=2$ this gives a new bound of $p\le \tfrac{1}{2}+\tfrac{1}{\sqrt{2n}}$ confirming a conjecture by Imamichi and Raymond (AQIS'18). Our bound implies that the previously known analytical constructions of $(3,2,\tfrac{1}{2}+\tfrac{1}{\sqrt{6}})$- , $(4,2,\tfrac{1}{2}+\tfrac{1}{2\sqrt{2}})$- and $(6,2,\tfrac{1}{2}+\tfrac{1}{2\sqrt{3}})$-QRACs are optimal. 
To obtain our bound we investigate the geometry of quantum states in the Bloch vector representation and make use of a geometric interpretation of the fact that any two quantum states have a non-negative overlap. 
\\ \\
\end{abstract}

\section{Introduction}

A Quantum Random Access Code (QRAC) brings about an interesting example of quantum advantages that arise from relaxing the requirement that transmission of information must be faithful. The Holevo bound implies that $n$ quantum bits cannot faithfully transmit more information than $n$ classical bits \cite{Hol73}. However, if faithfulness is not required, it becomes possible to obtain quantum advantages as also mentioned in \cite{Raymond,liabtr2016improved}. 
The study of QRACs 
seeks to answer the fundamental question of how well we can retrieve information after having stored it in a physical system. The principle behind a QRAC has a very long history relative to the field of research. It appeared already in 1983 in \cite{Wiesner1983} by the name of ``conjugate coding" and it has even been said \cite{brassard2006brief} that the inception of these codes can be traced back to late 1960's or early 1970's.
\\ \\
A Random Access Code (RAC) is a protocol for the following communication task: A sender (Alice) has to encode $n$ classical bits into $m$ classical bits ($m<n$) and send these to a recipient (Bob) who should then be able to use the received information to retrieve any of the initial $n$ bits with a probability of at least $p > 1/2$. 
In the quantum version of this problem (QRAC), Alice encodes the $n$ classical bits into $m$ \emph{qubits}.
We denote such encoding-decoding schemes by a triple, $(n,m,p)$-(Q)RAC.  We will present a formal definition shortly. 
\\ \\
After their first introduction in 1983 in \cite{Wiesner1983}  these codes did not receive much attention until  they were rediscovered and popularized in late 1990's in \cite{ambainis1998dense,Ambainis2002}. Here the authors exhibit $(2,1,\tfrac{1}{2}+\tfrac{1}{2\sqrt{2}})$- and $(3,1,\tfrac{1}{2}+\tfrac{1}{2\sqrt{3}})$-QRACs with the latter being attributed to Chuang. The optimality of these original single qubit QRACs follows, for example, from Theorem 3 in \cite{ambainis2008quantum} which states that any $(n,1,p)$-QRAC satisfies $p \leq \tfrac{1}{2}+\tfrac{1}{2\sqrt{n}}$ even with shared randomness. In \cite{ambainis2008quantum} it is also shown that if Alice and Bob have access to shared randomness, then their \textit{worst case success probability} can effectively be lifted to their \textit{average success probability}
\\ \\
Certain general limitations of QRACs have been established. In \cite{hayashi2006} it is shown that an $(n,m,p)$-QRAC exists only if $n < 4^{m}$. In \cite{Iwama2007} this bound is shown to be tight in the sense that an $(n,m,p)$-QRAC exists if and only if $n \leq 4^{m}-1$. In addition, the authors obtain a similar result for RACs, namely that an $(n,m,p)$-RAC exists if and only if $n<2^m$. Furthermore, \cite{nayak1999optimal} shows that for any $(n,m,p)$-QRAC, we must have that $m \geq (1-H(p))n$ where $H(p)$ is the binary entropy function
\begin{equation}
    H(p)=-p\log_2(p)-(1-p) \log_2(1-p).
\end{equation}
This is known as the Nayak bound. 
\\ \\
Recently, in \cite{Raymond} numerical searches have been conducted about $(n,2,p)$-QRACs. The authors find improved QRACs compared to the prior state of the art \cite{liabtr2016improved}. Their findings lead them to conjecture that $p \leq \tfrac{1}{2}+ \tfrac{1}{\sqrt{2n}}$ in any $(n,2,p)$-QRAC since the bound is tight when $n \leq 6$. For $n>6$ their
numerics suggest that the optimal success probability $p$ drops below the upper bound and that curiously it becomes advantageous to use mixed encoding states.
In this paper we confirm the above conjecture by proving a more general statement. We show that the average success probability of any $(n,m,p)$-QRAC must obey 
\begin{equation}
    p_{\text{avg}} \leq \tfrac{1}{2}+\tfrac{1}{2}\sqrt{\tfrac{2^{m-1}}{n}}.
\label{eq:Bound}    
\end{equation}
This also implies that an $(n,m,p)$-QRAC with shared randomness satisfies
\begin{equation}
    p\leq \tfrac{1}{2}+\tfrac{1}{2}\sqrt{\tfrac{2^{m-1}}{n}}.
\end{equation}
The above bound turns out to be new only for the case when $m=2$ and $m=3$ with $n\geq 16$. For $m=1$ it reproduces the bound from \cite{ambainis2008quantum} and for $m\ge 4$ it exceeds the Nayak bound for all values of $n$.
To obtain our bound (\ref{eq:Bound}) we investigate the geometry of quantum states in the Bloch vector representation which we will return to after a preliminary section. We show how one can get surprisingly far-reaching information about this geometry by interpreting the simple fact that the overlap of quantum states is non-negative. 
\\ \\
In the discussion section we summarize our findings by specifying the particular vector constructions in $\mathbb{R}^{4^m-1}$ QRACs need to be characterized by in order to reach our bound. It turns out, that the requirements for such constructions are slightly stricter when considering the worst case success probability compared to the average success probability. This observation allows us to conclude that there must exist some $n_{\max}(m)< 4^m-1$ such that the optimal worst case success probability of any $(n > n_{\max}(m),m,p)$ is stricly less than $\tfrac{1}{2}+\tfrac{1}{2}\sqrt{\tfrac{2^{m-1}}{n}}$. When comparing with the numerics in \cite{Raymond} we see that it is reasonable to conjecture that $n_{\max}(2)$ is equal to 6.

\section{Preliminaries}

In the following we denote the set of $N$-by-$N$ linear operators on $\mathbb{C}^N$ by $M_N(\mathbb{C})$. The set of $N$-level quantum states is then the following subset of $M_N (\mathbb{C})$
\begin{equation}
    Q_N=\{ \rho \in M_N(\mathbb{C}) \mid \rho \succeq 0   , \ \trace[\rho]=1    \}.
\end{equation}
Also, we denote the set of Hermitian, $N$-by-$N$ operators on $\mathbb{C}^N$ with trace equal to one by $H_N^1(\mathbb{C})$. Given $n\in \mathbb{N}$, we use $[n]$ to denote the set $\{1, ... ,n \}$.

\subsection{Definition of a QRAC}

The initial description of a QRAC in the introduction suggests that in order for Alice and Bob to achieve an $(n,m,p)$-QRAC they need a \textit{strategy} consisting of two parts: First, they need an encoding map 
\begin{equation}
    e: \{0,1 \}^n \longrightarrow Q_{2^m}
\end{equation}
that takes an input bit-string of length $n$ and outputs $m$ qubits, $e(x)=\rho_x$. Secondly, they need
a set of $n$ two-outcome POVMs denoted $\{D_i^0, D_i^1 \}_{i=1}^n$ such that for any $x \in \{0,1 \}^n$ we have 
\begin{equation}
    p \leq \trace[\rho_x D_i^{x_i}] =: p_{i,x}.
\end{equation}
A strategy thus defined is called a \textit{pure strategy}, which is synonymous with the term QRAC. If we allow cooperation between Alice and Bob in the sense that we give them access to a shared random string in order to agree on some joint strategy we get a \textit{QRAC with shared randomness}. This is defined as a probability distribution over pure stategies.
\\ \\
One can have different figures of merit when assessing the performance of a QRAC. The one we will mostly be concerned with is the worst case success probability. Hence we seek to upper bound the quantity 
\begin{equation}
    p:=\text{min}_{i,x} p_{i,x}.
\end{equation}
When proving the main result we will, in fact, upper bound the performance of a QRAC with respect to the average success probability,
\begin{equation}\label{aveprob}
   p_{\text{avg}}= \tfrac{1}{n2^n}\sum_{x \in \{0,1 \}^n} \sum_{i\in [n]} p_{i,x}.
\end{equation}
When we use the notation $(n,m,p)$-QRAC, we refer to the worst case success probability. And when we use the notation $(n,m)$-QRAC we refer to a QRAC with the usual parameters $n$ and $m$ but with a figure of merit specified in the particular context.
\\ \\
As mentioned in the introduction, it is shown in \cite{ambainis2008quantum} that an upper bound on the average success probability of a QRAC effectively becomes an upper bound on the worst case success probability of a QRAC with shared randomness.
\\ \\
Note, that the sum in (\ref{aveprob}) is invariant upon the interchange $p_{i,x}\rightarrow p_{i,\overline{x}}=\trace[\rho_{\overline{x}}D_i^{\overline{x_i}}]$ allowing us to rewrite (\ref{aveprob}) to get
\begin{equation}\label{aveprob1}
    p_{\text{avg}}=\tfrac{1}{n2^n}\sum_{x \in \{0,1 \}^n} \sum_{i\in [n]}\tfrac{1}{2}\big(\trace[\rho_xD_i^{x_i}]+\trace[\rho_{\overline{x}}D_i^{\overline{x_i}}]\big).
\end{equation}
It turns out, as mentioned earlier, that considering quantum states in the Bloch vector representation is useful. We will therefore review this in the following.

\subsection{Bloch Vector Representation of Quantum States}

\begin{definition}[Set of generators]
A subset $\{\vecsig_i \}_{i=1}^{N^2-1}$ of $M_N(\mathbb{C})$ is called a \textit{set of generators} if the following is fulfilled,
\begin{equation}
    \vecsig_j^{\dagger}=\vecsig_j, \ \trace[\vecsig_j]=0, \ \trace[\vecsig_j \vecsig_{j'}]=2\delta_{jj'}
\end{equation}
 for all $j, j' \in \{1, ... , N^2-1 \}$.
\end{definition}
A set of generators together with the identity operator forms an orthogonal basis for $M_N(\mathbb{C})$ with respect to the Hilbert-Schmidt inner product. Thus, if $\sigma$ is a set of generators, then for any $A \in M_N(\mathbb{C})$, there exists a unique vector of coefficients, $\boldsymbol{v}_{\sigma}=(v_{\sigma}^{1}, ... ,v_{\sigma}^{N^2}) \in \mathbb{C}^{N^2}$ such that
\begin{equation}
    A=v_{\sigma}^{N^2}I+ \sum_{i=1}^{N^2-1} v_{\sigma}^{i} \vecsig_i.
\end{equation}
Specifically, $\boldsymbol{v}_{\sigma}$ is given by
\begin{equation}
    (\tfrac{1}{2}\trace[A\vecsig_1], ... , \tfrac{1}{2}\trace[A\vecsig_{N^2-1}], \tfrac{1}{N}\trace[A]).
\end{equation}
Notice, that if $A \in H_N^1 (\mathbb{C})$ then the last entry of $v$ is equal to $\tfrac{1}{N}$ and the rest of the entries are real numbers. Hence the map, $\brho_{\sigma}: \mathbb{R}^{N^2-1} \rightarrow H_N^1(\mathbb{C})$ given by 
\begin{equation}\label{mapbrho}
\boldsymbol{r}=(r_1,...,r_{N^2-1}) \longmapsto \tfrac{1}{N}I+\tfrac{1}{2}\sum_{i =1}^{N^2-1} r_i  \vecsig_i   
\end{equation}
is a bijection with the inverse map, $\bbeta_{\sigma}: H_N^1(\mathbb{C}) \rightarrow \mathbb{R}^{N^2-1}$, given by 
\begin{align}
  \Lambda  \longmapsto (\trace[\vecsig_1 \Lambda] , ... , \trace[\vecsig_{N^2-1}\Lambda]).
\end{align}
This suggests the following definition:
\begin{definition}
Let $\rho \in Q_N$ and $\sigma$ be a set of generators. Then $\bbeta_{\sigma}(\rho)$ is called the \textit{Bloch vector representation of the quantum state $\rho$ with respect to $\sigma$}.
\end{definition}
Since $Q_N$ is a proper subset of $H_N^1(\mathbb{C})$, the set of Bloch vectors, $\bbeta_{\sigma}(Q_N)$, is a proper subset of $\mathbb{R}^{N^2-1}$. In the remainder of this paper we will refer to $\bbeta_{\sigma}(Q_N)$ as \textit{Bloch space}. Notice the following.
\begin{observation} \label{obsorth}
Let $\Lambda \in H_N^1 (\mathbb{C})$ and $\sigma$  and $\lambda$ be different sets of generators. There exists an orthogonal operator, $O_{\sigma \lambda} $ such that $\bbeta_{\sigma}(\Lambda) = O_{\sigma \lambda} \bbeta_{\lambda}(\Lambda)$.
\end{observation}
\begin{proof}
It is easily checked (by using the defining properties) that there exists $\{ \boldsymbol{v}^i\}_{i =1}^{N^2-1} \subset \mathbb{R}^{N^2-1}$ such that
\begin{equation}
     \vecsig_i = \sum_{j =1}^{N^2-1}  v^i_j  \veclam_j.
\end{equation}
fulfilling $\braket{\boldsymbol{v}^i, \boldsymbol{v}^{j}}=\delta_{ij}$. The $i$th coordinate of $\bbeta_{\sigma}   (\Lambda)$ can then be rewritten as
\begin{equation}
 \bbeta_{\sigma}   (\Lambda)_i= \sum_{j=1 }^{N^2-1} v^i_j \trace[ \veclam_j \Lambda] = \sum_{j =1}^{N^2-1} v^i_j \bbeta_{\lambda}(\Lambda)_j,
\end{equation}
which can equivalently be stated as the vector equation, $\bbeta_{\sigma}   (\Lambda)=O_{\sigma \lambda} \bbeta_{\lambda}   (\Lambda)$, if we define $O_{\sigma \lambda}$ as the $N^2-1$-by-$N^2-1$ operator with $v_j^i$ in its $(i,j)$th entry. The operator $O_{\sigma \lambda}$ is orthogonal since its rows are real, orthogonal unit vectors.
\end{proof}
Because of Observation \ref{obsorth}, all the following results about the geometry of $\bbeta_{\sigma}(Q_N)$ will be independent of the choice of generators. Therefore, we omit the subscript specifying the choice of generators. We will use $\partial \bbeta(Q_N)$ to denote the boundary of Bloch space. An element of the boundary corresponds to a density matrix that has at least one eigenvalue equal to zero making it infinitesimally close to an invalid state.

\section{Geometry of Bloch space} \label{Geomsection}

We start this section with a lemma that provides a useful characterization of Bloch space. Although the characterization has been given before \cite{Bengtsson_2012} we utilize it here to provide some intuition about the geometry of Bloch space. 
It will often be useful to note the following relation between the overlap of density matrices and the inner product of the corresponding Bloch vectors:
\begin{equation} \label{overlapbetweenqstates}
    \trace[\brho(\beta)\brho(\beta')] = \tfrac{1}{N}+\tfrac{1}{2}\braket{\beta,\beta'}.
\end{equation}

\begin{lemma} \label{hyperplanelemma}
For $\beta \in \bbeta(Q_N)$, let $\mathcal{A}_{\beta}$ be the following half-space,
\begin{equation}
    \mathcal{A}_{\beta}=\{z \in \mathbb{R}^{N^2-1} \mid \braket{z,\beta}\geq -\tfrac{2}{N}  \}.
\end{equation}
Then we have
\begin{equation}
    \bbeta(Q_N)=\bigcap_{\beta \in \bbeta(Q_N)} \mathcal{A}_{\beta}.
\end{equation}
\end{lemma}
\begin{proof}
First, let $\beta,\beta' \in \bbeta(Q_N)$. The overlap between $\brho(\beta)$ and $\brho(\beta')$ is given by (\ref{overlapbetweenqstates})
which must be non-negative. Hence, $\beta' \in \mathcal{A}_{\beta}$ which shows that
$\bbeta(Q_N) \subseteq \bigcap_{\beta \in \bbeta(Q_N)} \mathcal{A}_{\beta}$.
To show the other inclusion, we argue that if $z\nin\bbeta(Q_N)$, then $z\nin \bigcap_{\beta \in \bbeta(Q_N)} \mathcal{A}_{\beta}$. Assume that $z\nin\bbeta(Q_N)$. Even though $\brho(z)$ is not a density matrix, it is nevertheless a Hermitian matrix of trace one. Hence, $\brho(z)$ is not positive and we can find a quantum state $\ket{\psi}$ such that 
$\trace[\brho(z) \ket{\psi}\bra{\psi}]<0$.
By Eq.~(\ref{overlapbetweenqstates}), this implies that $z\not\in\mathcal{A}_{\bbeta(\ket{\psi}\bra{\psi})}$ allowing us to conclude that $z\not\in \bigcap_{\beta \in \bbeta(Q_N)} \mathcal{A}_{\beta}$, as desired.
\end{proof}
\begin{figure}
    \centering
\begin{tikzpicture}[thick, scale=.8]
\path[draw,fill, opacity=.1]
    (-1,-.75) rectangle (4.6,5.38);
    \filldraw[white] (2,2) circle (2.44949);
\draw[->] (2-3,2) -- (2+3,2) node [anchor=south] {$u_1$};
\draw[->] (2,2-2.9) -- (2,2+3.2);
\filldraw[black] (2,2) circle (1pt);
\draw (2,2) circle (2.44949);
\draw (2,2) circle (0.816497);
\draw[->] (2,2) -- (2+1.15*.8,2+1.15*0.8) node [anchor=north] {\textcolor{green}{\scriptsize{$\beta_1$}}};
\draw[->] (2,2) -- (2-1.6*1.38,2+0.75*1.38);
\draw (2-1.6*1.3,2+0.75*1.38) node [anchor=north] {\textcolor{blue}{\scriptsize{$\beta_2$}}};
\draw[->] (2,2) -- (2.134231,1.19461);



\draw (2,5) node[anchor=west] {$u_2$};
\draw (2.134231-.25,1.19461-.18) node[anchor=west] {\textcolor{red}{\scriptsize{$\beta_3$}}};

\path[draw,fill,opacity=0.3] (-3+2,1.26087+2) -- (1+2,-2.73913+2)--(-3+2,-2.73913+2)--cycle;
\path[draw,fill,opacity=0.3] (2+2.6,5.4) -- (1.96603-.4,-2.73913+2)--(4.6,-2.73913+2.0)--cycle;
\path[draw,fill,opacity=0.3] (-3+2,2.83333+1.15) -- (4.6,3.76667+1.15)--(4.6,5.37)-- (-1,5.37 ) --cycle;
\draw (2-0.402694-.7,2+2.41616+.5) node[anchor=west] {\textcolor{red}{\scriptsize{$\mathcal{A}_{\beta_3}^{C}$}}};
\filldraw[black] (2+1.02482,2-0.480384-.2) circle (0pt) node[anchor=west] {\textcolor{blue}{\scriptsize{$\mathcal{A}_{\beta_2}^{C}$}}};
\draw (2-0.869565*1.9,2-0.869565*1.4) node [anchor=west] {\textcolor{green}{\scriptsize{$\mathcal{A}_{\beta_1}^C$}}};

\end{tikzpicture}
    \caption{Bloch space intersected with the plane spanned by two orthogonal, unit vectors $u_1$ and $u_2$. $\beta_1$, $\beta_2$ and $\beta_3$ are Bloch vectors given by linear combinations of $u_1$ and $u_2$. The compliments of the half-spaces given by these Bloch vectors, in the sense of Lemma \ref{hyperplanelemma}, are shown. The grey areas are inaccesible for Bloch vectors according to Lemma \ref{hyperplanelemma} and the white area contains Bloch space.  The small and the large circle have radius $r_N$ and $R_N$, respectively. The distance from $\partial \mathcal{A}_{\beta_i}$ to the origin is $\tfrac{r_N R_N}{\norm{\beta_i}}$. $\beta_2$ is a pure state and its corresponding hyperplane is therefore tangent to the insphere. $\beta_3$ belongs to the insphere and its corresponding hyperplane is therefore tangent to the outsphere.}
    \label{hyperplanefig}
\end{figure}
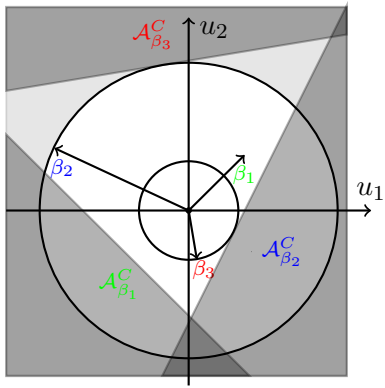

The boundary of the set $\mathcal{A}_{\beta}$ is a hyperplane with unit normal vector equal to $\beta/\norm{\beta}$ and a distance of $2/(N\norm{\beta})$ to the origin. The geometric interpretation of Lemma \ref{hyperplanelemma} is therefore that any vector in $\mathbb{R}^{N^2-1}$ is a valid Bloch vector if and only if, for all $\beta \in  \bbeta(Q_N)$, it lies in the half-space $\mathcal{A}_{\beta}$. See Figure \ref{hyperplanefig}.
\\ \\
Despite its seeming simplicity, Lemma~\ref{hyperplanelemma} allows us to easily obtain many interesting observations. We demonstrate this in the coming few paragraphs where we review some well-known facts about the Bloch space geometry that will be useful to us later on.
\\ \\
The well-known fact that Bloch space is a closed and convex subset of $\mathbb{R}^{N^2-1}$ follows by definition since it is the intersection of closed half-spaces. From Eq.~(\ref{overlapbetweenqstates}) and the fact that $\trace[\rho^2]\leq 1$, we see that the smallest ball $B_{R_N} \subseteq \mathbb{R}^{N^2-1}$ centered at the origin that contains Bloch space has radius 
\begin{equation}
 R_N = \sqrt{2\tfrac{N-1}{N}}.   
\end{equation}
Moreover, a state is pure if and only if its Bloch vector has norm $R_N$. From Lemma \ref{hyperplanelemma} we then get the well-known fact that the largest ball, centered at the origin, that is contained in Bloch space has radius 
\begin{equation}
 r_N=\tfrac{2}{NR_N}=\sqrt{\tfrac{2}{N(N-1)}}.
\end{equation}
Hence the following inclusions hold,
\begin{equation} \label{blochballs}
    B_{r_N} \subseteq \bbeta(Q_N) \subseteq B_{R_N}.
\end{equation}
The spheres of radii $r_N$ and $R_N$ centered at the origin will be refered to as the \textit{insphere} and the \textit{outsphere}, respectively. The inclusions in (\ref{blochballs}) coincide in case $N=2$ and Bloch space becomes the well-known three dimensional ball where orthogonal pure states correspond to antipodal points on the sphere of radius one.
\\ \\
Notice also, that a Bloch vector, $\beta \in \bbeta(Q_N)$, belongs to the boundary of Bloch space if and only if its corresponding density operator is not full rank. Hence, there exists a quantum state which is orthogonal to it. Therefore, $\partial \mathcal{A_{\beta}}\cap \bbeta(Q_N)$ is non-empty if and only if $\beta \in \partial \bbeta(Q_N)$. Let $u \in \mathbb{R}^{N^2-1}$ be a unit vector. The vector, $ur_N$, belongs to the surface of Bloch space if and only if $\partial \mathcal{A}_{ur_N}\cap \bbeta(Q_N)$ is non-empty. This means that $-uR_N \in \partial \bbeta(Q_N)$ since this is the only possible element of the set $\partial \mathcal{A}_{ur_N}\cap \bbeta(Q_N)$.  From this follows that boundary points in a distance of $r_N$ and $R_N$ from the origin come in dual pairs which is also shown in \cite{Kimuraspherical} using more elaborate spectral arguments.
\\ \\
Lemma \ref{hyperplanelemma} can also be used to prove the following Lemma which to the best of our knowledge is hitherto unknown. Although the proof is very simple, Lemma \ref{hyperplanelemma} is, in fact, the crucial observation that makes the proof of our main theorem possible.

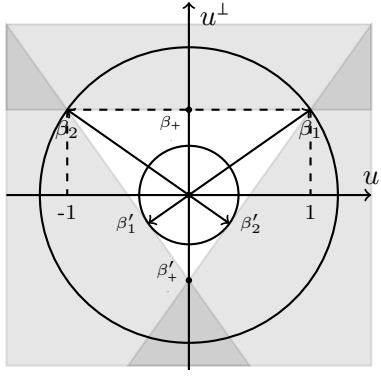
\begin{figure}[t]
    \centering
\begin{tikzpicture}[thick, scale=.8]
\draw[->] (2-3,2) -- (2+3,2) node [anchor=south] {$u$};
\draw[->] (2,2-2.9) -- (2,2+3.2);
\filldraw[black] (2,2) circle (1pt);
\draw (2,2) circle (2.44949);
\draw (2,2) circle (0.816497);
\draw[->] (2,2) -- (4,3.41421) node [anchor=north] {\scriptsize{$\beta_1$}};
\draw[->] (2,2) -- (0, 3.41421) node [anchor=north] {\scriptsize{$\beta_2$}};
\draw[->] (2,2) -- (1.33333,1.5286) node [anchor=east] {\tiny{$\beta_1'$}};
\draw[->] (2,2) -- (2.66667, 1.5286) node [anchor = west] {\tiny{$\beta_2'$}};
\filldraw[black] (1.7,2.9) circle (0pt) node[anchor=south] {\tiny{$\beta_{+}$}};
\filldraw[black] (1.65,.4) circle (0pt) node[anchor=south] {\tiny{$\beta'_+$}};
\filldraw[black] (2,3.41421) circle (1pt) node[anchor=south] {};
\filldraw[black] (2,2-1.41421) circle (1pt) node[anchor=south] {};
\draw (2,5) node[anchor=west] {$u^{\perp}$};
\draw[dashed] (4,3.41421) --  (4,2) node [anchor=north] {\scriptsize{1}};
\draw[dashed] (0,3.41421) --  (0,2) node [anchor=north] {\scriptsize{-1}};
\draw[dashed] (0,3.41421) --  (4,3.41421) node [anchor=south] {};
\path[draw,fill, opacity=0.1] (-1,4.82843) -- (3,-0.828427)--(-1,-0.828427)--cycle;
\path[draw,fill, opacity=0.1] (1,-0.828427) -- (5,4.82843)--(5,-0.828427)--cycle;
\path[draw,fill, opacity=0.1] (-1,3.41421) -- (5,3.41421)--(5,4.82843)--(-1,4.82843)--cycle;
\end{tikzpicture}
    \caption{Two section of $\mathbb{R}^{N^2-1}$ along $u=(\beta_1 - \beta_2)/2$ and $u^{\perp}=(\beta_1+\beta_2)(2\sqrt{R_N^2-1})^{-1}$ where $\bt_1$ and $\bt_2$ are orthogonal pure states and $\beta_{+}=\tfrac{1}{2} (\beta_1+\beta_2)$ and $\beta'_+=-\tfrac{r_NR_N}{R_N^2-1}\beta_+$. Due to convexity and the fact that all the points $\beta_1$, $\beta_2$, $\beta_1'$, $\beta_2'$, $\beta_+$ and $\beta_+'$ belong to the boundary we see that the white area \textit{is exactly} Bloch space in this particular two-section.
    }
    \label{uppercompfig}
\end{figure}

\begin{lemma} \label{lemmauppercomp}
For any two $\beta_1 , \beta_2 \in \bbeta(Q_N)$ we have $\tfrac{1}{2}\norm{\beta_1-\beta_2} \leq 1$.
\end{lemma}
\begin{proof}
We see that 
\begin{equation} 
 \tfrac{1}{2}\norm{\beta_1- \beta_2}=\tfrac{1}{2}\sqrt{\norm{\beta_1}^2+\norm{\beta_2}^2-2\braket{\beta_1,\beta_2}} .
\end{equation}
We have $\norm{\bt_1}, \norm{\bt_2}\leq R_N$ and from Lemma \ref{hyperplanelemma} we know that $\braket{\beta_1,\beta_2} \geq -r_NR_N$. Hence, we obtain
\begin{equation}
    \tfrac{1}{2}\norm{\bt_1-\bt_2}\leq \tfrac{1}{2} \sqrt{2R_N^2+2r_NR_N}=1.
\end{equation}
\end{proof}
Notice, that the upper bound of this Lemma can only be saturated for a pair of Bloch vectors corresponding to two orthogonal pure states. For such a pair, $\beta_1$, $\beta_2$, we have that $\beta_+:=\tfrac{1}{2}(\beta_1+\beta_2)$ (orthogonal with $\tfrac{1}{2}(\beta_1-\beta_2)$ since $\norm{\beta_1}=\norm{\beta_2}$) is a valid Bloch vector with norm equal to $\sqrt{R_N^2-1}$, since it is a convex combination of valid Bloch vectors. For $N=2$ we have $\bt_+=0$ and for $N>2$ we have $\bt_+\in \partial \bbeta(Q_N)$ since it corresponds to a convex combination of rank 1 operators. Hence its rank is strictly less than $N$.
\\ \\
In Figure \ref{uppercompfig} we show the intersection of the plane spanned by $u$ parallel with $\beta_1-\beta_2$ and $u^{\perp}$ parallel with $\beta_1+\beta_2$. The reason why Bloch space in this two-section looks exactly like this is given by the following observation which is a modification of a Theorem in \cite{Kimuraspherical}. The proof of the exact statement used here is included in Appendix A for completion.

\begin{restatable}{observation}{obsrestat} \label{obs1}
Let $\bt \in \bbeta(Q_N) \backslash \{ 0\}$. Then 
\begin{equation}
 \tfrac{\bt}{1-N\lambda_{\max}}, \tfrac{\bt}{1-N\lambda_{\min}} \in \partial \bbeta(Q_N)  
\end{equation}
where $\lambda_{\min}$ and $\lambda_{\max}$ is the smallest and largest eigenvalue of $\brho(\bt)$, respectively. In particular, $-\beta \in \partial \bbeta(Q_N)$ if and only $\lambda_{\max}= r_N R_N$.
\end{restatable}

For two orthogonal pure states, $\beta_1$ and $\bt_2$ the eigenvalues of $\beta_+$ are $\tfrac{1}{2}$ (with multiplicity 2) and $0$ (with multiplicity $N-2$). Thus for $N>2$, Observation \ref{obs1} gives us that
\begin{equation}
 \beta'_+:=\tfrac{\beta_+}{1-N/2}=-\tfrac{r_NR_N}{R_N^2-1}\beta_+ \in \partial \bbeta(Q_N).  
\end{equation}
This, by Lemma \ref{hyperplanelemma}, belongs to both $\partial \mathcal{A}_{\beta_1} $ and $  \partial \mathcal{A}_{\beta_2}$.

\section{Upper Bounds on QRACs}

In this section we utilize the characterization of the geometry of Bloch space given in the previous section in the context of QRACs. We set $N=2^m$. First observe the following Lemma.
\begin{lemma} \label{lemmapovm}
Let $\{D^{0} , D^{1} \}$ be a POVM and let,
\begin{equation}\label{povm1}
    D^{x}=(x+(-1)^x\alpha_0) I+(-1)^x\sum_{i=1}^{4^m-1}\alpha_i \vecsig_i
\end{equation}
for $x\in \{0,1 \}$ be an expansion in terms of a set of generators. Then $0 \leq \alpha_{0} \leq 1$ and 
\begin{equation}\label{constraintPOVM}
 \norm{\boldsymbol{\alpha}} \leq \tfrac{1}{2\sqrt{r_{2^m}R_{2^m}}},
\end{equation}
with equality only if $\alpha_{0}=\tfrac{1}{2}$.
\end{lemma}
\begin{proof}
Since $0 \preceq D^0 \preceq I$, we get that $0\le \trace{D^0}\le \trace(I)$. Combining this with $\trace(D^0) = \alpha_{0}\trace(I)$ gives that $0 \leq \alpha_{0} \leq 1$.
Now, if we let $z_1 =\tfrac{r_{2^m}R_{2^m}}{\alpha_{0}}\boldsymbol{\alpha}$ and $z_2= -\tfrac{\alpha_{0}}{1-\alpha_{0}}z_1$ then the POVM elements can be factorized as 
\begin{equation}
 D^0=2^m \alpha_{0}\brho(z_1)    \ , \ D^1=2^m(1-\alpha_{0})\brho(z_2).
\end{equation}
Since we require that $D^0, D^1 \succeq 0$ we need $z_1,z_2 \in \bbeta (Q_{N})$. A necessary condition for this is given by Lemma \ref{hyperplanelemma} in that $z_1 \in \mathcal{A}_{z_2}$ which, after some rearranging, gives the restriction
\begin{equation} \label{intermediate3}
    \norm{\boldsymbol{\alpha}}^2 \leq \tfrac{\alpha_{0}(1-\alpha_{0})}{r_{2^m}R_{2^m}} .
\end{equation}
The right hand side, when viewed as a function of $\alpha_{0}$, is maximized if $\alpha_{0}=1/2$. In that case (\ref{intermediate3}) gives the desired bound from (\ref{constraintPOVM}).
\end{proof}
If the bound in (\ref{constraintPOVM}) is saturated then 
\begin{equation} \label{POVMvectors}
 D^{x}=\tfrac{1}{r_{2^m}R_{2^m}} \brho ((-1)^{x}\sqrt{r_{2^m}R_{2^m}}\boldsymbol{\hat{\alpha}}),
\end{equation}
where $\boldsymbol{\hat{\alpha}} = \boldsymbol{\alpha}/\norm{\boldsymbol{\alpha}}$.
Now, notice the following:
\begin{observation} \label{obs3}
Let $u \in \mathbb{R}^{4^m-1}$ be a unit vector. Suppose $\pm \sqrt{r_{2^m}R_{2^m}}u \in \partial \bbeta(Q_{2^m})$ and let $\rho_+=\brho(\sqrt{r_{2^m} R_{2^m}}u) $ and  $\rho_-=\brho(-\sqrt{r_{2^m} R_{2^m}}u)$. We then have $\rho_+ + \rho_- = r_{2^m} R_{2^m} I$ and $\trace[\rho_+\rho-]=0$. 
\end{observation}
\begin{proof}
This follows immediately from the definition of the map $\brho$ in (\ref{mapbrho}) and (\ref{overlapbetweenqstates}).
\end{proof}
This observation implies that a POVM that saturates the bound in (\ref{constraintPOVM}), corresponds to a projective measurement.
\\ \\
The following Lemma is a modification and extension of Lemma 5 in \cite{ambainis2008quantum}. In Appendix B we include a proof of the exact statement used here for completion.



\begin{restatable}{lemma}{lemrestat} \label{mancinskalemma}
For any set of vectors, $\{\mu_i \}_{i=1}^n\subseteq \mathbb{R}^{N}$, the inequality
\begin{equation} \label{upperboundonunitvectors}
    \sum_{x \in \{0,1 \}^n}\norm{\sum_{i\in [n]} (-1)^{x_i} \mu_i} \leq  2^n \sqrt{\sum_{i\in [n]} \norm{\mu_i}^2} 
\end{equation}
holds with equality if and only if the $\mu_i$'s are orthogonal.
\end{restatable}

The preceding three Lemmas can be combined in the following Theorem which is the main result of this work.
\begin{theorem} \label{maintheorem}
The average success probability of any $(n,m)$-QRAC is upper bounded as
\begin{equation} \label{maxprob}
    p_{\text{avg}} \leq \tfrac{1}{2}+\tfrac{1}{2}\sqrt{\tfrac{2^{m-1}}{n}}.
\end{equation}
\end{theorem}
\begin{proof}
Consider an $(n,m)$-QRAC with POVMs given as in (\ref{povm1}). The Bloch vector for encoding the string $x \in \{ 0,1\}^n$ is denoted $\beta_x$. Thus, we find 
\begin{equation}
   \trace[D_i^{x_i} \brho(\beta_x)]+\trace[D_i^{\overline{x}} \brho(\beta_{\overline{x}})]=1+(-1)^{x_i}\big(\braket{\alpha^{(i)} , \beta_x} - \braket{\alpha^{(i)} , \beta_{\overline{x}} }\big).
\end{equation}
Using this in (\ref{aveprob1}) we obtain
\begin{equation} \label{intermediateIV}
    p_{\text{avg}}=\tfrac{1}{2}+\tfrac{1}{n2^n}\sum_{x \in \{0,1 \}^n} \sum_{i \in [n]} (-1)^{x_i} \tfrac{1}{2}\braket{\alpha^{(i)} , \beta_x-\beta_{\overline{x}}}.
\end{equation}
Now, let $T_x:=\sum_{i \in [n]} (-1)^{x_i} \alpha^{(i)}$. Then, owing to the three last Lemmas, we can make the following estimate on (\ref{intermediateIV}):
\begin{multline}
    p_{\text{avg}}=\tfrac{1}{2}+\tfrac{1}{n2^n}\sum_{x \in \{ 0,1\}^n}\tfrac{1}{2} \braket{T_x , \beta_x-\beta_{\overline{x}}} \leq \tfrac{1}{2}+\tfrac{1}{n2^n} \sum_{x \in \{0,1 \}^n} \tfrac{1}{2}\norm{T_x}\norm{\bt_x-\bt_{\overline{x}}} \leq \tfrac{1}{2}+\tfrac{1}{n2^n} \sum_{x \in \{0,1 \}^n} \norm{T_x}  \\ \leq \tfrac{1}{2}+\tfrac{1}{n} \sqrt{\sum_{i\in [n]}\norm{\alpha^{(i)}}^2}  \leq \tfrac{1}{2}+\tfrac{1}{2} \sqrt{\tfrac{1}{nr_{2^m}R_{2^m}} }=\tfrac{1}{2}+\tfrac{1}{2}\sqrt{\tfrac{2^{m-1}}{n}}.
\end{multline}
where we used Lemma \ref{lemmauppercomp} in the second inequality, Lemma \ref{mancinskalemma} in the third inequality and Lemma \ref{lemmapovm} in the fourth.
\end{proof}
In analogy with Theorem 3 in \cite{ambainis2008quantum} we therefore have the following corollary:
\begin{corollary}
Any $(n,m,p)$-QRAC with shared randomness fulfills
\begin{equation}
    p\leq \tfrac{1}{2}+\tfrac{1}{2}\sqrt{\tfrac{2^{m-1}}{n}}.
\end{equation}
\end{corollary}

\section{Discussion}

For $m=2$ as well as $m=3$ with $n \geq 16$, Theorem \ref{maintheorem} gives a new and tighter upper bound on the average success probability when compared to Nayak's bound. 
It also provides a tighter upper bound of
\begin{equation}
    p\leq \tfrac{1}{2}+\tfrac{1}{2}\sqrt{\tfrac{2^{m-1}}{n}}
\label{eq:Ubound}
\end{equation}
on the worst case success probability $p$. Our upper bound (\ref{eq:Ubound}) coincides with the success probabilities of the $(3,2,\tfrac{1}{2}+\tfrac{1}{\sqrt{6}})$-QRAC found in \cite{Raymond} and the $(4,2,\tfrac{1}{2}+\tfrac{1}{2\sqrt{2}})$- and $(6,2,\tfrac{1}{2}+\tfrac{1}{2\sqrt{3}})$-QRACs one can construct by concatenation of the known, optimal $(2,1,\tfrac{1}{2}+\tfrac{1}{2\sqrt{2}})$- and $(3,1,\tfrac{1}{2}+\tfrac{1}{2\sqrt{3}})$-QRACs. This shows that these analytically constructed QRACs are optimal. In addition, our upper bound (\ref{eq:Ubound}) matches the success probability of the $(5,2)$-QRAC found by the numerical procedure put forth in~\cite{Raymond}.
\\ \\
When $n>6$, this numerical procedure finds $(n,2)$-QRACs with success probabilities strictly below our bound~(\ref{eq:Ubound}) leading us to believe that for $m=2$ our bound ceases to be tight at $n=7$.
Additionally, in view of the specific Bloch vector configuration which is necessary to reach our bound~(\ref{eq:Ubound}), it is possible to show the following (for a proof see Appendix C), 
\begin{restatable}{corollary}{correstat}\label{CorDis}
For $m>1$ there exists $n_{\max}(m) \le 4^m-1$ such that any $(n\ge n_{\max}(m),m,p)$-QRAC fulfills \begin{equation}
 p<\tfrac{1}{2}+\tfrac{1}{2}\sqrt{\tfrac{2^{m-1}}{n}}.   
\end{equation}
\end{restatable}
The numerics carried out in 
\cite{Raymond} suggest that $n_{\max}(2)=7$.
The fact that optimal $(4,2,\tfrac{1}{2}+\tfrac{1}{2\sqrt{2}})$- and $(6,2,\tfrac{1}{2}+\tfrac{1}{2\sqrt{3}})$-QRACs 
can be obtained by taking two copies of the $(2,1,\tfrac{1}{2}+\tfrac{1}{2\sqrt{2}})$- and $(3,1,\tfrac{1}{2}+\tfrac{1}{2\sqrt{3}})$-QRACs respectively makes us wonder if it is possible to identify natural conditions under which concatenation of optimal QRACs gives rise to a QRAC that remains optimal.
In general concatentation of QRACs does not preserve optimality. Indeed, for $n_1\neq n_2$ it is easy to find examples where 
concatenating optimal $(n_1,m_1,p_1)$- and $(n_2,m_2,p_2)$-QRAC yields a sub-optimal $(n_1+n_2, m_1+m_2)$-QRAC. Yet perhaps this sub-optimality results from the fact that the success probability of the concatenated code is the minimum of $p_1$ and $p_2$. In view of this, we would like to pose the following question: Given an optimal $(n,m,p)$-QRAC, does taking $k$-copies of it produce an optimal $(kn,km,p)$-QRAC for all $k \in \mathbb{N}$? One could also restrict this question to the special case where we take $k$ copies of the optimal $(2,1)$- or $(3,1)$-QRAC. In this case $k$-fold concatenation produces codes with
parameters $(2k,k,\tfrac{1}{2}+\tfrac{1}{2\sqrt{2}})$ and $(3k,k,\tfrac{1}{2}+\tfrac{1}{2\sqrt{3}})$ respectively. In both of these cases, the worst case success probability takes the form
\begin{equation}\label{conj}
    p=\tfrac{1}{2}+\tfrac{1}{2}\sqrt{\tfrac{m}{n}}.
\end{equation}
In fact, one can show that 
\begin{equation}
    1-H\left(\tfrac{1}{2}+\tfrac{1}{2}\sqrt{\tfrac{m}{n}} \right) \leq \tfrac{m}{n}
\end{equation}
for all $m< n \in \mathbb{N}$ and thus the expression in Eq.~(\ref{conj}) lies below Nayak's bound.
Hence, it would be interesting to know whether the worst case success probability of any $(n,m,p)$-QRAC must be upper bounded by $\tfrac{1}{2}+\tfrac{1}{2}\sqrt{\tfrac{m}{n}}$.



\section*{Acknowledgements}
This paper is based on S.~Storgaard's bachelor's thesis.
L.~Mančinska acknowledges support by Villum Fonden via the QMATH Centre of Excellence (Grant No. 10059) and Villum Young Investigator grant (No. 37532).
\section*{Data availability}
Data sharing not applicable to this article as no datasets were generated or analysed during the current study.
\bibliographystyle{bibstyle}
\bibliography{bibliography.bib}

\newpage

\section*{Appendix A: Proof of Observation \ref{obs1}}
\obsrestat*

\begin{proof}
We list the eigenvalues of $\brho(\beta)$ as $\lambda_{\text{max}} \geq \cdots \geq \lambda_{\text{min}} \geq 0$. Then the eigenvalues of $\tfrac{1}{2} \sum_{i=1}^{N^2-1} \bt_i  \vecsig_i$ can be listed as
\begin{equation}
\lambda_{\max}-\tfrac{1}{N} \geq \cdots \geq \lambda_{\min} - \tfrac{1}{N}.    
\end{equation}
Let $\bt'=\gamma \bt$ for some $\gamma \in \reals$. If $\gamma \geq 0$ The eigenvalues of $\brho(\gamma \beta)$ can be listed as
\begin{equation}
    \tfrac{1}{N}(1-\gamma)+\gamma \lambda_{\max} \geq \cdots \geq \tfrac{1}{N}(1-\gamma)+\gamma \lambda_{\min} \geq 0
\end{equation}
and hence $\bt' \in \partial \bbeta (Q_N)$ is in the boundary of Bloch space if $\gamma=\tfrac{1}{1-N\lambda_{\min}}$. Similarly if $\gamma <0$ then $\bt' \in \partial \bbeta (Q_N)$ if $\gamma =\tfrac{1}{1-N\lambda_{\max}}$.

\end{proof}

\section*{Appendix B: Proof of Lemma \ref{mancinskalemma}} \label{App}
\lemrestat*
\begin{proof}

The inequality in (\ref{upperboundonunitvectors}) holds if all the $\mu_i$'s are 0, so we assume they are not all 0. We can interpret the sum on the left hand side of (\ref{upperboundonunitvectors}) as the inner product of 
\begin{equation}
 y_1=(1, ... ,1) \in \mathbb{R}^{\{0,1 \}^n} \cong \mathbb{R}^{2^n}   
\end{equation}
and a vector $y_2 \in \mathbb{R}^{\{0,1 \}^n}$, whose entry corresponding to bit string $x\in \mathbb{R}^{\{0,1 \}^n}$ is given by
\begin{equation} \label{y2comp}
    (y_2)_x=\norm{\sum_{i\in [n]}(-1)^{x_{i}} \mu_{i}}.
\end{equation} 
Applying the Cauchy-Schwarz inequality, we get that the left hand side of (\ref{upperboundonunitvectors}) is upper bounded by 
\begin{equation} \label{int2}
    \norm{y_1}\norm{y_2}=
    \sqrt{2^n}\sqrt{\sum_{x \in \{0,1 \}^n}\norm{\sum_{i\in [n]} (-1)^{x_i} \mu_i}^2}.
\end{equation} 
We claim now that
\begin{equation} \label{int}
    \sum_{x \in \{0,1 \}^n}\norm{\sum_{i\in [n]} (-1)^{x_i} \mu_i}^2=2^n \sum_{i \in [n]} \norm{\mu_i}^2.
\end{equation}
Maybe we can refer to Lemma 5 in \cite{ambainis2008quantum}. Afterall the proofs are essentially the same... This can be proved by induction on $n$. First, Eq.~(\ref{int}) holds for $n=1$ since 
\begin{equation}
   \norm{\mu_1}^2+\norm{-\mu_1}^2=2\norm{\mu_1}^2.
\end{equation}
Assume that Eq.~(\ref{int}) holds for $n=k$ and consider the case when $n=k+1$. By explicitly carrying out the sum over $x_{k+1} \in \{0,1 \}$ on the left hand side of Eq.~(\ref{int}) we get
\begin{equation}\label{intermediate4}
\sum_{x \in \{0,1 \}^{k}} \Big[ \norm{  \big( (-1)^{x_1} \mu_1 + ... + (-1)^{x_k} \mu_k \big) + \mu_{k+1} }^2 
 + \norm{  \big( (-1)^{x_1} \mu_1 + ... + (-1)^{x_k} \mu_k \big) - \mu_{k+1} }^2 \Big].
\end{equation}
Applying the parallelogram identity, i.e. 
\begin{equation}
 \norm{u_1+u_2}^2+ \norm{u_1-u_2}^2=2(\norm{u_1}^2+\norm{u_2}^2)  ,
\end{equation}
the expression in (\ref{intermediate4}) equals
\begin{equation}
   2 \sum_{x \in \{0,1 \}^{k}}\big( \norm{ \textstyle \sum_{i \in [k]} (-1)^{x_i} \mu_i }^2+\norm{\mu_{k+1}}^2  \big).
\end{equation}
Finally, applying the induction hypothesis, we complete the inductive step as follows:
\begin{equation}
    2\big( 2^k \sum_{i \in [k]} \norm{\mu_i}^2+2^k \norm{\mu_{k+1}}^2 \big)=2^{k+1}\sum_{i \in [k+1]}\norm{\mu_{i}}^2.
\end{equation}
Now, by inserting (\ref{int}) in (\ref{int2}) we can conclude (\ref{upperboundonunitvectors}). 
\\ \\
Recall that $\braket{y_1 ,y_2} = \norm{y_1}\norm{y_2}$ if and only if $y_2 =k y_1$ for some $k\in\mathbb{R}$. Hence, the bound in (\ref{upperboundonunitvectors}) holds with equality if and only if the quantity $(y_2)_x$ in Eq.~(\ref{y2comp}) is equal to some constant $c$ independent of $x\in\{0,1\}^n$. In other words, for all $x \in \{0,1 \}^n$ we must have that
\begin{equation}
    \sum_{i\neq j \in [n]} (-1)^{x_i+x_j}\braket{\mu_i ,\mu_j}=c^2-\sum_{i\in [n]} \norm{\mu_i}^2,
\label{eq:const}    
\end{equation}
where the right hand side is constant. Now, fix $m\in[n]$. The left hand side of (\ref{eq:const}) can be rewritten as
\begin{equation}
    \sum_{i\neq j \in [n]\backslash \{ m\}}(-1)^{x_i+x_j}\braket{\mu_i,\mu_j} +2\sum_{i\in [n]\backslash \{ m\}}(-1)^{x_m+x_i}\braket{\mu_m,\mu_i}
\end{equation}
This must be invariant upon the interchange $x_m \rightarrow \overline{x_m}$. Since $(-1)^{x_m}=-(-1)^{\overline{x_m}}$ we have that 
\begin{equation} \label{eq.const2}
    \sum_{i\in [n]\backslash \{ m\}}(-1)^{x_m+x_i}\braket{\mu_m,\mu_i}=0.
\end{equation}
Now fix $m'\in [n]\backslash \{m \}$ and rewrite (\ref{eq.const2}) as
\begin{equation}
    (-1)^{x_m+x_{m'}}\braket{\mu_m,\mu_{m'}}+\sum_{i\in [n]\backslash \{ m,m'\}}(-1)^{x_m+x_i}\braket{\mu_m,\mu_i}=0.
\end{equation}
This must be invariant upon the interchange $x_{m'}\rightarrow \overline{x_{m'}}$ so we get $\braket{\mu_m,\mu_{m'}}=0$. This completes the proof.
\end{proof}

\section*{Appendix C: Proof of Corollary \ref{CorDis}} \label{AppC}
\correstat*
\begin{proof}
The average success probability of an $(n,m)$-QRAC reaches the bound in Theorem \ref{maintheorem} if and only if it is given by the following Bloch vector configuration:
\begin{enumerate}
    \item A set $\{\nu_i\}_{i=1}^n$ of orthogonal unit vectors such that $\pm \sqrt{r_{2^m}R_{2^m}}\nu_i \in \partial \bbeta(Q_{2^m})$. The POVM for measuring the $j$th bit is then associated with the pair $\pm \sqrt{r_{2^m}R_{2^m}}\nu_j$ in the sense of (\ref{POVMvectors}).
    \item For the encodings, $2^{n-1}$ pairs of orthogonal pure state Bloch vectors, $\{\beta_x, \beta_{\overline{x}} \}$, where
    \begin{equation}\label{2}
        \tfrac{1}{2}(\beta_x-\beta_{\overline{x}})=\tfrac{1}{\sqrt{n}}\sum_{i\in [n]} (-1)^{x_i} \nu_i=: V_x,
    \end{equation}
with $\{\nu_i\}_{i=1}^n$ given as in 1.  
    \end{enumerate}
Let $V:=\text{span} \big\{V_x \mid x\in \{ 0,1\}^n  \big\}$. Assume, for some $(n,m)$-QRAC, that 1 and 2 above are fulfilled. This implies that its average success probability is $\tfrac{1}{2}+\tfrac{1}{2}\sqrt{\tfrac{2^{m-1}}{n}}$.
In view of Eqs.~(\ref{overlapbetweenqstates}), (\ref{POVMvectors}) and (\ref{2}) one can use the decomposition 
\begin{equation}
\beta_{x,\overline{x}}= \pm \tfrac{1}{2}(\beta_x-\beta_{\overline{x}})+\tfrac{1}{2}(\beta_x+\beta_{\overline{x}}),
\end{equation}
to calculate the probability of correctly decoding the $i$th bit of $x$ to be $x_i$ as
\begin{align}
    p_{i,x}&=\tfrac{1}{r_{2^m}R_{2^m}}\trace\left[\brho\left(\tfrac{1}{2}(\beta_x-\beta_{\overline{x}})+\tfrac{1}{2}(\beta_x+\beta_{\overline{x}})\right)  \brho \left(\sqrt{r_{2^m}R_{2^m}}\nu_i\right) \right]
    \\&=\tfrac{1}{r_{2^m}R_{2^m}}\left(\tfrac{1}{2^{m}}+\tfrac{\sqrt{r_{2^m}R_{2^m}}}{2}\Big\langle \tfrac{1}{2}(\beta_x-\beta_{\overline{x}})+\tfrac{1}{2}(\beta_x+\beta_{\overline{x}}),\nu_i\Big\rangle \right)
    \\&=\tfrac{1}{2}+\tfrac{1}{2}\tfrac{1}{\sqrt{r_{2^m}R_{2^m}}}\left(\tfrac{1}{\sqrt{n}}+\Big\langle\tfrac{1}{2}(\beta_x+\beta_{\overline{x}}),\nu_i\Big\rangle\right)
    \\ &= \tfrac{1}{2}+\tfrac{1}{2}\sqrt{\tfrac{2^{m-1}}{n}}\big( 1 +\sqrt{n}\Big\langle\tfrac{1}{2}(\beta_x+\beta_{\overline{x}}),\nu_i\Big\rangle \big).
\end{align}
As noted, the average of this is $\tfrac{1}{2}+\tfrac{1}{2}\sqrt{\tfrac{2^{m-1}}{n}}$. It follows that
the worst case success probability reaches $\tfrac{1}{2}+\tfrac{1}{2}\sqrt{\tfrac{2^{m-1}}{n}}$ only if, we additionally have that for all $x\in \{0,1 \}^n$, $\tfrac{1}{2}(\beta_x+\beta_{\overline{x}})\in V^{\perp}$. Unless $m=1$ we have that $\tfrac{1}{2}(\beta_x+\beta_{\overline{x}})\neq 0$ i.e.~for $m>1$, $V^{\perp}$ must be of non-vanishing dimension. From this we conclude the desired.
\end{proof}

\end{document}